\documentstyle[12pt]{article}
\def\be{\begin{eqnarray}}
\def\ee{\end{eqnarray}}
\def\ba{\begin{array}}
\def\ea{\end{array}}
\def\p{\phi}

\def\a{\alpha}

\def\pa{\partial}

\def\n{\nabla}

\def\G{{\cal G}}
\def\B{{\cal B}}
\def\A{{\cal A}}
\def\X{{\cal X}}
\def\H{{\cal H}}
\def\S{{\cal S}}

\def\L{{\cal L}}
\def\E{{\cal E}}

\def\T{{\cal T}}

\def\5{^{(5)}}
\def\D{^{(D)}}

\begin{document}

\begin{center}
{\Large Nonlinear hidden symmetries in General Relativity and String\\
\vskip 0.25cm
 Theory: a matrix generalization of the Ernst potentials}
\end{center}
\vskip 1cm
\begin{center}
{\bf \large {Nandinii Barbosa--Cendejas$^\ddagger$\footnote{E-mail:
nandinii@correo.fie.umich.mx}, Alfredo
Herrera--Aguilar$^\natural$\footnote{E-mail:
alfredo.herrera.aguilar@gmail.com},}}
\end{center}
\begin{center}
{\bf \large {Konstantinos Kanakoglou$^\natural$$^*$\footnote{E-mail:
kanakoglou@ifm.umich.mx, kanakoglou@hotmail.com} and Joannis E.
Paschalis$^*$\footnote{E-mail: paschalis@physics.auth.gr}}}
\end{center}
\vskip 0.1cm
$^\ddagger$Facultad de Ingenier\'{\i}a El\'ectrica,
Universidad Michoacana de San Nicol\'as de Hidalgo. \\
Edificio B, Ciudad Universitaria, C.P. 58040, Morelia,
Michoac\'{a}n, M\'{e}xico. \\
$^\natural$Instituto de F\'{\i}sica y Matem\'{a}ticas, Universidad
Michoacana de San Nicol\'as de Hidalgo. \\
Edificio C--3, Ciudad Universitaria, C.P. 58040, Morelia,
Michoac\'{a}n, M\'{e}xico. \\
$^*$School of Physics, Nuclear and Elementary Particle Physics Department, \\
Aristotle University of Thessaloniki (AUTH), 54124 Thessaloniki,
Greece.


\begin{abstract}
In this paper we recall a simple formulation of the stationary
electrovacuum theory in terms of the famous complex Ernst
potentials, a pair of functions which allows one to generate new
exact solutions from known ones by means of the so--called nonlinear
hidden symmetries of Lie--B\"acklund type. This formalism turned out
to be very useful to perform a complete classification of all 4D
solutions which present two spacetime symmetries or possess two
Killing vectors. Curiously enough, the Ernst formalism can be
extended and applied to stationary General Relativity as well as the
effective heterotic string theory reduced down to three spatial
dimensions by means of a (real) matrix generalization of the Ernst
potentials. Thus, in this theory one can also make use of nonlinear
matrix hidden symmetries in order to generate new exact solutions
from seed ones. Due to the explicit independence of the matrix Ernst
potential formalism of the original theory (prior to dimensional
reduction) on the dimension $D$, in the case when the theory
initially has $D\ge 5$, one can generate new solutions like {\it
charged} black holes, black rings and black Saturns, among others,
starting from uncharged field configurations.
\end{abstract}


\section{Introduction}

It is well known that both, the stationary action and the coupled
field equations of the Einstein--Maxwell theory can be formulated in
terms of a pair of very simple complex functions that were called
{\it Ernst potentials} after their inventor \cite{e,iw}. In the
language of these potentials, the black holes of Schwarzschild and
Kerr, Reissner--Nordstr\"on and Kerr--Newmann adopt a very simple
form, as well as some cosmological models, among other exact
solutions \cite{e,kramer}. Indeed, this formalism facilitates the
general study of the symmetries of the theory and, hence, the
construction of new exact solutions by means of very well-known
solution--generating techniques (see, for instance,
\cite{kinnersleyetal}).

It turns out that the Ernst formalism can be generalized to
low--energy effective string theories and General Relativity with
extra dimensions in terms of matrix potentials instead of complex
functions (see \cite{kramer},\cite{ggk}--\cite{belver}, for
instance). This matrix formalism also enables one to study the
complete symmetry group of the underlying theory and to apply
generalized solution--generating techniques with matrix charges
involved \cite{CSEMDA}--\cite{hk5}. In particular, this matrix
formalism can be applied to the classification and construction of
charged black holes, black rings and black Saturns in 5D and
multiple black rings in $D\ge 6$ in the framework of such theories
\cite{emparanetal}--\cite{elvangfigueras}.

In this paper we first recall the derivation of the Ernst potentials
for the stationary Einstein--Maxwell theory and write both field
equations and the effective action in their language. We further
refer to the stationary formulation of the low--energy heterotic
string theory, and the corresponding field equations, in terms of a
pair of matrix Ernst potentials that closely resembles the
formulation of the stationary theory of electrovacuum in the
language of the complex Ernst potentials. A fact that, in principle,
allows one to generalize all the so far obtained results in the
stationary Einstein--Maxwell theory to the realm of the stationary
heterotic string theory.

As an extra bonus, within the framework of higher dimensional
General Relativity and the low energy limit of heterotic string
theory, the matrix Ernst potentials can be used to classify and
construct exact solutions that corresponds to higher dimensional
objects like black holes, black rings, black Saturns and multiple
black rings. A sketch of how this program can be performed is given
at the end of this paper.

\section{Ernst potentials in the stationary Einstein--Maxwell theory}

In this section we briefly review the derivation of the Ernst
potentials within the framework of the stationary Einstein--Maxwell
theory basically following the work given by \cite{iw}.

Let us consider the 4D action of the electrovacuum theory \be
\S_{EM}= \int d^4x\mid
G\mid^{\frac{1}{2}}\left(^4\!R-\frac{1}{4}F^2_{mn}\right), \ee where
$G$ is the determinant of the metric $G_{mn}$, $F_{mn}=\pa_mA_{n} -
\pa_nA_{m}$, $A_{m}$ is the gauge field, $^4\!R$ is the scalar
curvature in 4D and $m,n,=0,1,2,3;$ \ $\mu, \nu=1,2,3.$

Consider now the stationary ansatz for the metric \be
ds^2=G_{mn}dx^mdx^n=-f(dt+\omega_{\mu}dx^{\mu})^2+
f^{-1}\gamma_{\mu\nu}dx^{\mu}dx^{\nu}, 
\ee where $f$, $\gamma_{\mu\nu}$ and $\omega_{\mu}$ are quantities
independent on $t$.

Indices of spatial coordinates are raised and lowered with the aid
of the metric tensor $\gamma_{\mu\nu}$ and its inverse
$\gamma^{\mu\nu}$, unless otherwise indicated through a left
superindex $^{(0)}$.

Thus, if $F_{mn}$ is a covariant tensor, then
$$F^{\a\beta}=\gamma^{\a\mu}\gamma^{\beta\nu}F_{\mu\nu}\quad \mbox{\rm and} \quad
^{(0)}\!F^{\a\beta}= g^{\a m}g^{\beta n}F_{mn}.$$

The three--dimensional vector $\omega_{\mu}$ can always be dualized
through an invariant torsion vector in the following form \be
f^{-2}\tau^{\mu}=-\gamma^{-1/2}\epsilon^{\mu\rho\sigma}
\partial_{\rho}\omega_{\sigma}
\label{dual} \ee or, equivalently, \be
f^{-2}\vec\tau=-\nabla\times\vec\omega, \ee
by making use of the
three--dimensional vectorial calculus which employs
$\gamma_{\mu\nu}dx^{\mu}dx^{\nu}$ as background metric.

Let us now consider a stationary electromagnetic field $F_{mn}=\pa_m
A_n-\pa_n A_m$ with the given metric.

The stationarity condition $\pa_0 A_m=0$ for the electric field
implies \be F_{0\nu}=-\pa_{\nu}A_0, \label{elec} \ee while the
sourceless Maxwell equations \be
\pa_{\nu}\left[(-g)^{1/2}\,^{(0)}F^{m\nu}\right]=0 \label{max} \ee
in the case when $m=\mu$ provide us with the magnetic components \be
^{(0)}\!F^{\mu\nu}=f\gamma^{-1/2}\epsilon^{\mu\nu\rho}\pa_{\rho}\psi,
\label{mag} \ee in terms of the scalar magnetic potential $\psi$.

It turns out that all the remaining components can be expressed as
functions of these six magnitudes; for instance, \be
^{(0)}\!F^{0\nu}=\omega_{\mu}^{(0)}\!F^{\mu\nu}+\gamma^{\mu\nu}F_{0\mu},
\label{id} \ee is an identity that is directly inferred from the
stationary metric.

By substituting the relations (\ref{id}), (\ref{mag}), (\ref{elec})
and (\ref{dual}) in the Maxwell equations (\ref{max}) with $m=0$ one
gets
\be \n\left(f^{-1}\n A_0\right)=-f^{-2}\vec\tau\cdot\n\psi.
\label{m=0} \ee

By rewriting $F_{\mu\nu}\equiv\pa_{\mu}A_{\nu}-\pa_{\nu}A_{\mu}$
with the aid of the relations (\ref{elec}) and (\ref{mag}), and
making use of the expression for the cyclic identity
$\epsilon^{\mu\nu\rho}\pa_{\rho}F_{\mu\nu}=0,$ one obtains
\be
\n\left(f^{-1}\n \psi\right)=f^{-2}\vec\tau\cdot\n A_0.
\label{idcic} \ee

Now one is able to introduce the scalar complex potential \be
\Phi=A_0+i\psi, \label{peme} \ee which is precisely the
electromagnetic Ernst potential.

By combining (\ref{m=0}) and (\ref{idcic}) one obtains a single
complex equation \be \n\left(f^{-1}\n
\Phi\right)=if^{-2}\vec\tau\cdot\n \Phi. \label{ecmax} \ee

Thus, in this way we have reduced the stationary Maxwell equations
to a single equation in terms of the complex electromagnetic Ernst
potential.

On the other side, within the framework of the Einstein equations
for the gravitational field, it turns out convenient to express the
Ricci tensor \be R_{mn}=
\pa_m\Gamma^a_{na}-\pa_a\Gamma^a_{mn}+\Gamma^a_{bm}\Gamma^b_{an}-
\Gamma^a_{ba}\Gamma^b_{mn} \ee in terms of a complex
three--dimensional vector $\vec G$ defined by \be 2f\vec G=\n
f+i\vec\tau \label{2fg} \ee for the general case of the stationary
metric.

In this way we can obtain the following relations \be
-f^{-2}R_{00}=\n\vec G+\left(\vec G^*-\vec G\right)\cdot \vec G,
\label{r00} \ee \be
-2if^{-2}\,^{(0)}\!R_0^{\mu}=\gamma^{-1/2}\epsilon^{\mu\rho\sigma}
\left(\pa_{\sigma}G_{\rho}+G_{\rho}G^*_{\sigma}\right), \label{r0mu}
\ee \be
f^{-2}\!\left(\!\gamma_{\rho\mu}\gamma_{\sigma\nu}\,^{(0)}\!R^{\mu\nu}\!-\!
\gamma_{\rho\sigma}R_{00}\!\right)\!\!=\!\!R_{\rho\sigma}(\gamma)
\!+\!G_{\rho}G^*_{\sigma}\!+\!G^*_{\rho}G_{\sigma}, \label{rmn} \ee
where $R_{\rho\sigma}(\gamma)$ stands for the Ricci tensor
calculated through the three--dimensional metric
$\gamma_{\mu\nu}dx^{\mu}dx^{\nu}$.

Thus, from the above obtained formulas, for the energy--momentum
tensor of the electromagnetic field \be -4\pi
T_{mn}=g^{ab}F_{ma}F_{nb}-\frac{1}{4}g_{mn}F_{ab}F^{ab} \ee one gets
the following relations \be
\frac{1}{2}F_{mn}F^{mn}=\left(\n\psi\right)^2- \left(\n
A_0\right)^2, \ee \be 8\pi
f^{-1}T_{00}=\left(\n\psi\right)^2+\left(\n A_0\right)^2,
\label{t00} \ee \be 4\pi
f^{-1}\,^{(0)}\!T^{\mu}_0=\gamma^{-1/2}\epsilon^{\nu\rho\sigma}
\left(\pa_{\rho}\psi\right)\left(\pa_{\sigma}A_0\right), \label{t0m}
\ee \be -4\pi f^{-1}\,^{(0)}T^{\mu\nu}=
\left(\pa^{\mu}\psi\right)\left(\pa^{\nu}\psi\right)+
\left(\pa^{\mu}A_0\right)\left(\pa^{\nu}A_0\right) -\frac{1}{2}\
\gamma^{\mu\nu} \left[\left(\n\psi\right)^2\!+\!\left(\n
A_0\right)^2\right], \label{tmn} \ee where
$\pa^{\mu}=\gamma^{\mu\nu}\pa_{\nu}$.

By making use of the Einstein equations \be R_{mn}=-8\pi T_{mn}, \ee
from the relations (\ref{r0mu}) and (\ref{t0m}) one obtains \be
\n\times\vec\tau=-4\n\psi\!\times\n A_0=
i\n\times\left(\Phi\n\Phi^*-\Phi^*\n\Phi\right). \ee

In this way, the following equation \be
\vec\tau+i\left(\Phi^*\n\Phi-\Phi\n\Phi^*\right)=\n\chi \label{xi}
\ee defines the scalar potential $\chi$ up to an additive constant.

Now let us define the complex scalar potential \be
E=f-\Phi\Phi^*+i\chi, \label{gep} \ee called gravitational Ernst
potential.

This potential allows one to obtain, from the relations (\ref{2fg})
and (\ref{xi}), the following equality
\be f\vec G=\frac{1}{2}\n
E+\Phi^*\n\Phi. \label{fG} \ee

By substituting (\ref{fG}) in the gravitational field equations
(\ref{r00}) and (\ref{t00}), and making use of the Maxwell equations
(\ref{ecmax}), we obtain a single equation \be f\n^2E=\left(\n
E+2\Phi^*\n\Phi\right)\cdot\n E; \label{gp} \ee on the other hand,
the relation (\ref{ecmax}) can be expressed in the following way:
\be f\n^2\Phi=\left(\n E+2\Phi^*\n\Phi\right)\cdot\n\Phi.
\label{emp} \ee

It is evident that from the definition (\ref{gep}), one can obtain
the following expression for the function $f:$
\be
f=\frac{1}{2}\left(E+E^*\right)+\Phi\Phi^*. \label{f} \ee

Thus, relations (\ref{gp}) and (\ref{emp}) are the well--known
differential Ernst equations for the stationary electrovacuum.

Finally, the gravitational field equations (\ref{rmn}) and
(\ref{tmn}) reduce to the following expression \be
-f^2R_{\mu\nu}=\frac{1}{2}E,_{(\mu}E^*_{,\nu)}+\Phi
E,_{(\mu}\Phi^*_{,\nu)}+\Phi^*E^*,_{(\mu}\Phi_{,\nu)}-
\left(E+E^*\right)\Phi,_{(\mu}\Phi^*_{,\nu)}, \label{einstein} \ee
where the symmetrization of indices are defined in the following
form \be 2E,_{(\mu}E^*_{,\nu)}\equiv(\pa_{\mu}E)(\pa_{\nu}E^*)+
(\pa_{\nu}E)(\pa_{\nu}E^*). \ee

In this way, the field equations for the Ernst potentials (\ref{gp})
and (\ref{emp}), together with the Einstein equations
(\ref{einstein}), determine the dynamics of the field system of the
stationary Einstein--Maxwell theory.

This system of self--consistent second order differential equations,
despite their apparent simplicity, has no general solution at the
moment. Only particular solutions are known in the literature and it
is of great relevance to obtain new solutions possessing a coherent
and consistent physical interpretation. It is worth noticing that
precisely at this point is where the solution--generating techniques
(which make use of nonlinear hidden symmetries to construct new
solutions starting from seed ones) can be of great help towards this
aim.

\subsection{Effective action of the stationary EM theory and Ernst potentials}

Now let us express the effective action of the stationary
Einstein--Maxwell theory from which one can derive both the Einstein
equations (\ref{einstein}), and the Ernst equations (\ref{gp}) and
(\ref{emp}) by the variational method.

By redefining the electromagnetic Ernst potential as follows \be
\Phi\equiv\frac{1}{\sqrt2}F, \ee the effective stationary action of
the Einstein--Maxwell theory adopts the following form \be
^4\S_{EM}= \int d^3x\mid
g\mid^{\frac{1}{2}}\left(-^3\!R+^3\!\L_{EM}\right), \nonumber \ee
where the matter Lagrangian $^3\!\L_{EM}$ is given by \be ^3\L_{EM}=
\frac{1}{2f^2}\!\left|\nabla E\!+F^*\!\n F\right|^2\!-
\frac{1}{f}\left|\nabla F\right|, \ee where now
$f=\frac{1}{2}(E+E^*+FF^*)$. It is a straightforward exercise to
vary this action and obtain the above quoted Einstein and Ernst
equations.

\section{Low energy effective action of heterotic string and matrix Ernst potentials}

The effective action of the low--energy limit of the heterotic
string at tree level takes into account just the massless modes of
the theory and possesses the form \cite{ms,s}
\be \S\D=\int d\D
x\mid G\D\mid^{\frac{1}{2}}e^{-\p\D}\left(R\D+
\p\D_{;M}\p^{(D);M}\right.- \nonumber \ee \be
\left.\frac{1}{12}\!H\D_{MNP}H^{(D)MNP}-
\frac{1}{4}F^{(D)I}_{MN}F^{(D)IMN}\right), \label{accion} \ee where
\be F^{(D)I}_{MN}=\pa_MA^{(D)I}_N-\pa _NA^{(D)I}_M, \qquad \qquad
I=1,\,2,\,...,n; \nonumber \ee \be
H\D_{MNP}=\pa_MB\D_{NP}-\frac{1}{2}A^{(D)I}_M\,F^{(D)I}_{NP}+
\mbox{{\rm cyclic perms. of}\,M,\,N\, and \,P.} \nonumber \ee Here
$G\D_{MN}$ is the metric, $B\D_{MN}$ is the anti--symmetric
Kalb--Ramond tensor field, $\p\D$ is the dilaton and $A^{(D)I}_M$ is
a set of $U(1)$ vector fields ($I=1,\,2,\,...,n$). $D$ is the
dimensionality of the spacetime and $M,N,P=1,2,3,...,10$. In the
consistent critical case (where the quantum theory is free of
anomalies) $D=10$ and $n=16$, but we shall leave these parameters
arbitrary in our analysis for the sake of generality.

By following Maharana and Schwarz \cite {ms}, and Sen \cite {s}, we
further perform the dimensional reduction of this model on a
$D-3=d$--torus. Thus, the resulting three--dimensional, stationary
theory possesses the $SO(d+1,d+1+n)$ symmetry group and describes
gravity in terms of the metric tensor \be
g_{\mu\nu}\!=\!e^{-2\p}\!\left(G\D_{\mu\nu}\!-
\!G\D_{p+3,\mu}G\D_{q+3,\nu}G^{pq}\right), \label{g3D} \ee where the
subscripts $p,q=1,2,...,d$; coupled to the following set of
three--dimensional fields:

a) scalar fields \be G=\left (G_{pq}= G\D_{p+3,q+3}\right ),\qquad
B=\left ( B_{pq}=B\D_{p+3,q+3}\right), \nonumber \ee \be A=\left (
A^I_p=A^{(D)I}_{p+3}\right ) ,\qquad \p=\p\D-\frac{1}{2}{\rm ln \,|
det}\,G|. \label{scalars} \ee

b) antisymmetric tensor field of second rank
\be
B_{\mu\nu}=B\D_{\mu\nu}-4B_{pq}A^p_{\mu}A^q_{\nu}-
2\left(A^p_{\mu}A^{p+d}_{\nu}-A^p_{\nu}A^{p+d}_{\mu}\right), \ee
(hereafter we shall set $B_{\mu\nu}=0$ in order to remove the
effective three--dimensional cosmological constant from our
consideration).

c) vector fields $A^{(a)}_{\mu}=
\left((A_1)^p_{\mu},(A_2)^{p+d}_{\mu},(A_3)^{2d+I}_{\mu}\right)$
($a=1,...,2d+n$) \be
(A_1)^p_{\mu}\!=\!\frac{1}{2}G^{pq}G\D_{q+3,\mu},\quad
(A_3)^{I+2d}_{\mu}\!=\!-\frac{1}{2}A^{(D)I}_{\mu}\!+\!A^I_qA^q_{\mu},\nonumber
\ee \be
(A_2)^{p+d}_{\mu}\!=\!\frac{1}{2}B\D_{p+3,\mu}\!-\!B_{pq}A^q_{\mu}\!+\!
\frac{1}{2}A^I_{p}A^{I+2d}_{\mu}. \ee

In three dimensions all vector fields $A^{(a)}_{\mu},$ can be
dualized on--shell with the aid of the pseudoscalar potentials $u,$
$v$ and $s$ in the following form:
\begin{eqnarray}
\nabla\times\overrightarrow{A_1}&=&\frac{1}{2}e^{2\p}G^{-1}
\left(\nabla u+(B+\frac{1}{2}AA^T)\nabla v+A\nabla s\right),
\nonumber                          \\
\nabla\times\overrightarrow{A_3}&=&\frac{1}{2}e^{2\p} (\nabla
s+A^T\nabla v)+A^T\nabla\times\overrightarrow{A_1}, \nonumber
\\
\nabla\times\overrightarrow{A_2}&=&\frac{1}{2}e^{2\p}G\nabla v-
(B+\frac{1}{2}AA^T)\nabla\times\overrightarrow{A_1}+
A\nabla\times\overrightarrow{A_3}.
\end{eqnarray}
Thus, the resulting effective three--dimensional theory describes
the scalars $G$, $B$, $A$ and $\p$ and the pseudoscalars $u$, $v$
and $s$ coupled to the metric $g_{\mu\nu}$.

We further define the so--called {\it matrix} Ernst potentials (MEP)
from all these scalar and pseudoscalar potentials in order to
express the low--energy effective action of the heterotic string in
a similar form to the formulation of the stationary
Einstein--Maxwell theory in terms of the complex Ernst potentials
\cite{hk3}:
\be \X= \left( \ba{cc}
-e^{-2\p}+v^TXv+v^TAs+\frac{1}{2}s^Ts&v^TX-u^T \cr Xv+u+As&X \ea
\right) \quad {\rm and} \quad \A=\left( \ba{c} s^T+v^TA \cr A \ea
\right), \label{XA}\ee where $X=G+B+\frac{1}{2}AA^T$. These
potentials are of dimensions $(d+1) \times (d+1)$ and $(d+1) \times
n,$ respectively.

The physical meaning of their components are as follows: The
relevant information about the gravitational field is encoded in the
potential $X$, while its rotational nature is parameterized by the
pseudoscalar $u$; $\p$ is the dilatonic field; $v$ is related to the
multi--dimensional components of the antisymmetric tensor field of
Kalb--Ramond. Finally, $A$ and $s$ represent electric and magnetic
potentials.

\subsection{Stationary effective action of heterotic string and field
equations in the language of MEP}

In terms of MEP the effective three--dimensional theory adopts the
form \cite{hk3}: \be ^3\!\S=\int d^3x \mid
g\mid^{\frac{1}{2}}\{-^3\!R+^3\!\L_{{}_{HS}}\}, \label{ES3D}\ee
where the matter Lagrangian is given by \be \L_{{}_{HS}}= {\rm
Tr}\,\left[ \frac{1}{4}\left(\nabla\X-\nabla\A\A^T\right)\G^{-1}
\left(\nabla\X^T-\A\nabla\A^T\right)\G^{-1}+\frac{1}{2}\nabla\A^T\G^{-1}\nabla\A\right],
\label{matter3D}\ee $^3\!R$ is the three--dimensional curvature
scalar and the matrix potential $\X$ is defined by
$\X=\G+\B+\frac{1}{2}\A\A^T$.

The symmetric part of the potential is given by the matrix
$\G=\frac{1}{2}\left(\X+\X^T-\A\A^T\right)$ and the antisymmetric
one by $\B=\frac{1}{2}\left(\X-\X^T\right);$ these matrices are
parameterized as follows: \be \G= \left( \ba{cc}
-e^{-2\p}+v^TGv&v^TG \cr Gv&G \ea \right) \qquad \mbox{\rm and}
\qquad \B=\left( \ba{cc} 0&v^TB-u^T \cr Bv+u&B \ea \right). \ee

By making use of the conventional method of variations, from the
effective action (\ref{ES3D}) one obtains both the {\it Einstein
equations} \be ^3R_{\mu\nu}= {\rm
Tr}\left[\frac{1}{4}\left(\nabla_{\mu}\X\!-\!\nabla_{\mu}
\A\A^T\right)\!\G^{-1}
\!\left(\nabla_{\nu}\X^T\!-\!\A\nabla_{\nu}\A^T\right)\!\G^{-1}
+\frac{1}{2}\nabla_{\mu}\A^T\G^{-1}\nabla_{\nu}\A\right], \ee as
well as the {\it Ernst equations} for the potentials $\X$ and $\A$
which represent the matter sector of the theory: \be
\n^2\X-2\left(\n\X-\n\A\A^T\right)(\X+\X^T-\A\A^T)^{-1}\n\X=0,
\nonumber \ee \be
\n^2\A-2\left(\n\X-\n\A\A^T\right)(\X+\X^T-\A\A^T)^{-1}\n\A=0,
\nonumber \ee as a matrix version of the equations of the stationary
Einstein--Maxwell theory.

As we have pointed out above, these differential equations are not
so simple to solve in a closed form. However, one can make use of
the similarity which exists with respect to the equations of the
stationary Einstein--Maxwell theory in order to guess and write down
the solutions in a direct way or to perform nonlinear symmetries to
generate new exact solutions from known ones (for some examples see
\cite{HSsolns}).

\section{Heterotic string {\it vs.} Einstein--Maxwell}

Thus, it has been shown that there exists a close relation between
the stationary effective actions of the heterotic string and the
Einstein--Maxwell theory: \be \X\longleftrightarrow -E, \qquad
\qquad \A\longleftrightarrow F, \label{mapa}  \ee \be {\it
matrix\,\, transposition}\quad\longleftrightarrow\quad {\it
complex\,\, conjugation}. \nonumber \ee

One can realize that the relation (\ref{mapa}) allows us to
generalize in a straightforward way the results obtained within the
framework of the Einstein--Maxwell theory to the realm of the
heterotic string (where a suitable physical interpretation will be
needed since more fields are involved) by making use of the MEP
formalism. Actually, the four--dimensional Einstein--Maxwell theory,
being reduced to three dimensions, can be written as a special case
of the MEP formalism with some peculiarities in terms of the complex
Ernst potentials $E$ and $F$ \cite{hk5}.

Let us rewrite them in a less conventional form \be
-\X_{{}_{EM}}={\rm Re}E+\sigma_2\,{\rm Im}E, \quad \A_{{}_{EM}}={\rm
Re}F+\sigma_2\,{\rm Im}F, \quad {\rm where} \quad \sigma_2= \left(
\ba{cc} 0&-1 \cr 1&0 \ea \right). \label{xa}\ee We can treat these
matrices as the matrix Ernst potentials (\ref{XA}) of the $D=4$
theory (\ref{accion}) with $\phi^{(4)}=B^{(4)}_{MN}=0$. Then we
conclude that we need two Abelian gauge fields $n=2$ and that they
should satisfy the following constraint \be s^1=A^2={\rm Re}F, \quad
-s^2=A^1={\rm Im}F. \ee Note, that $s^I$ $(I=1,2)$ describe the
magnetic potentials, whereas $A^I$ are the electric ones. Thus, both
Maxwell fields arising in the framework of the representation
(\ref{XA})--(\ref{matter3D}) and (\ref{xa}) turn out to be mutually
conjugated (i.e. $F_{MN}^{(4)2}=\tilde F_{MN}^{(4)1}$ in four
dimensions). Next, for the single extra metric component one has:
\be G=-\frac{1}{2}\left(E+E^*+FF^*\right)\equiv f, \qquad {\rm and}
\qquad u={\rm Im}E. \label{f=G}\ee By taking into account that
$\G=G$, and by substituting equations (\ref{xa}) and (\ref{f=G})
into the matter Lagrangian (\ref{matter3D}), we obtain \be
\L_{{}_{EM}}=\frac{1}{2f^2}\left |\nabla E+F^*\nabla F\right |^2 -
f^{-1}\left|\nabla F\right|^2. \ee As we already have seen, this is
precisely the matter Lagrangian of the stationary Einstein--Maxwell
theory. Thus, our MEP formulation of the heterotic string theory
includes the Einstein--Maxwell theory as a special case.

It is worth noticing as well that the higher dimensional General
Relativity theory can also be written in terms of a matrix Ernst
potential when reduced to three dimensions. This fact corresponds to
a special case in which the matter degrees of freedom of the
low--energy heterotic string theory (\ref{accion}) vanish: the
anti--symmetric Kalb--Ramond tensor field $B\D_{MN}=0,$ the dilaton
$\p\D=0$ and the Abelian gauge fields $A^{(D)I}_M=0,$ so that the
matrix Ernst potential is symmetric $\X=\G$ and $\B=\A=0.$ It should
also be mentioned that the three-dimensional dilaton field must
remain nontrivial since it is identified with the determinant of the
extra dimensional metric according to the definitions (\ref{g3D})
and (\ref{scalars}).

Thus, this parametrization of the above mentioned higher dimensional
theories in terms of the MEP can be very useful when performing a
complete classification of the higher dimensional ($D\ge 5$) black
objects (holes, rings, Saturns, etc.) obtained in the literature
during last years (see \cite{emparanetal} for a review).

\section{Nonlinear hidden symmetries and their possible applications in $D\ge 5$}

One of the advantages of the (matrix) Ernst potential formalism is
that the study of symmetries (conservation laws) of the stationary
effective action can be performed in a very straightforward way. It
turns out that the complete symmetry group, apart from rescalings
and shifts of the Ernst potentials, involves nonlinear symmetries
that were initially called {\it hidden} in the framework of General
Relativity; moreover, an infinite--dimensional double hidden
symmetry structure was revealed for string effective actions
\cite{Gao}. In particular, these symmetries act nontrivially in the
charge space of a seed solution and can be used to generate new
charged solutions from uncharged ones. There also other effects when
applying this symmetries (see, for instance,
\cite{kinnersleyetal,ggk,hk5,hamf,haptv}.

Here we shall quote just the symmetries which preserve the
asymptotic properties of the (matrix) Ernst potentials for
physically meaningful field configurations of both the stationary
Einstein--Maxwell and low--energy heterotic string theories. These
symmetries possess the same form for both theories and allow one to
generate similar solutions in both realms \cite{hk5}.

For the stationary Einstein--Maxwell theory we have:
\be
E\rightarrow E, \qquad F\rightarrow e^{i\alpha}F; \qquad \qquad
\qquad \left({\rm EMT}\right) \ee
\be E\rightarrow \frac {E+i\epsilon}{1+i\epsilon E}, \qquad
F\rightarrow \frac {1-i\epsilon}{1+i\epsilon E}F; \qquad \qquad
\qquad \left ({\rm NET}\right) \ee
\be E\rightarrow \frac {E+\frac {1}{2}\left |\lambda_{\H}\right |^2
-\bar \lambda_{\H}F} {1-\bar \lambda_{\H}F+\frac {1}{2} \left
|\lambda_{\H}\right |^2E}, \qquad F\rightarrow \frac {\left (1+\frac
{1}{2}\left |\lambda_{\H}\right |^2\right )F- \lambda_{\H}\left (
E+1\right )} {1-\bar \lambda_{\H}F+\frac {1}{2} \left
|\lambda_{\H}\right |^2E}, \qquad \left({\rm NHT}\right) \ee where
EMT stands for Electric--Magnetic Transformation, NET for Normalized
Ehlers Transformation and NHT for Normalized Harrison
Transformation, the parameter $\lambda _{\H}$ is complex while the
parameters $\alpha$ and $\epsilon$ are real. It is easy to check
that when the parameters $\lambda _{\H},$ $\alpha$ and $\epsilon$
vanish, one recovers the original (seed) potentials.

On the other hand, for the stationary low--energy effective action
of the heterotic string we have the following matrix symmetries:
\be
\X\rightarrow\X+\lambda_{\X}, \qquad \A\rightarrow\A \qquad {\rm
with} \qquad \lambda_{\X}^T=-\lambda_{\X}
\ee
\be
\A\rightarrow\A+\lambda_{\A}, \qquad \X\rightarrow\X+
\A\lambda_{\A}^T+\frac{1}{2}\lambda_{\A}\lambda_{\A}^T
\ee
\be
\A\rightarrow\A\T, \qquad   \X\rightarrow\X, \qquad {\rm where}
\qquad \T\T^T=1
\ee
\be
\X\rightarrow\S^T\X\S, \qquad \A\rightarrow\S^T\A,  \qquad {\rm
with} \qquad  \S\rightarrow (\S^T)^{-1}.
\ee
\be &&\A\rightarrow \left(1+\Sigma\lambda_{\E}\right)
\left(1+\X\lambda_{\E}\right)^{-1}\A, \nonumber \qquad \qquad \qquad
\qquad \qquad {\rm (NET)}
\\
&&\X\rightarrow\left(1+\Sigma\lambda_{\E}\right)
\left(1+\X\lambda_{\E}\right)^{-1}\X(1-\lambda_{\E}\Sigma)+
\Sigma\lambda_{\E}\Sigma. \ee
\be
\A\rightarrow\left(1+\frac{1}{2}\Sigma\lambda_{\H}\lambda_{\H}^T\right)
\left(1-\A\lambda_{\H}^T+\frac{1}{2}\X\lambda_{\H}\lambda_{\H}^T\right)^{-1}\times\nonumber
\ee \be \left(A-\X\lambda_{\H}\right)+\Sigma\lambda_{\H}, \nonumber
\qquad \qquad \qquad \qquad {\rm (NHT)} \ee
\be
\X\rightarrow\left(1+\frac{1}{2}\Sigma\lambda_{\H}\lambda_{\H}^T\right)
\left(1-\A\lambda_{\H}^T+\frac{1}{2}\X\lambda_{\H}\lambda_{\H}^T\right)^{-1}\times\nonumber
\ee \be
\left[\X+\left(\A-\frac{1}{2}\X\lambda_{\H}\right)\lambda_{\H}^T\Sigma\right]
+\frac{1}{2}\Sigma\lambda_{\H}\lambda_{\H}^T\Sigma. \ee
where
$\lambda_{\E}^T=-\lambda_{\E}$ and $\lambda_{\H}$ is a real
rectangular matrix of dimension $(d+1)\times n.$

The last pair of nonlinear symmetries can be applied to construct
new exact solutions starting from known (sometimes quite simple)
field configurations in both theories. As an example one can cite
the construction of the of the Reissner--Nordstr\"om solution
starting from the Schwarzschild black hole one in the 4D
Einstein--Maxwell theory.

We finally quote a procedure to construct new charged field
configurations from known neutral solutions within the framework of
theories like General Relativity and the effective low--energy
action of the heterotic string with more than four dimensions (in
the spirit of \cite{hamf,haptv}). Thus, this procedure can be
applied to the construction of charged black holes, black rings and
black Saturns if $D=5$, and charged multiple black rings in $D=6$:

\begin{enumerate}
\item
Write the exact solution of the uncharged field configuration (black
ring or black Saturn, for instance) in the form of a generalized
Weyl metric \cite{emparanreallWeyl,harmark} by making use of a
suitable coordinate system.

\item
Identify the symmetric and antisymmetric parts of the matrix Ernst
potential $\X$.

\item
Perform the nonlinear hidden symmetry NHT on the matrix Ernst
potentials $\X$ and $\A$.

\item
Write the new higher--dimensional charged exact solution with the
aid of $\X$ and $\A$.

\item
Physically interpret the new solution with the aid of the behaviour
of the fields and their properties.

\end{enumerate}

This procedure can be performed also in a wider class of
higher--dimensional field configurations that have the form of a
stationary axisymmetric seed solution (the so--called
Weyl--Papapetrou class) \cite{othersolutions} and it is interesting
to see what kind of physical configurations arise after applying the
MEP symmetry method.

\section{Acknowledgements}
One of the authors (AHA) is really grateful to the organizers of the
Miniworkshop {\it Symmetries 2010}, held at UAZ, for providing a
friendly and warm environment for developing fruitful and
illuminating discussions. KK would like to thank the whole staff of IFM, UMSNH.
He acknowledges support from CONACYT 60060--J. This research was supported by grants
CIC--UMSNH--4.16 and CONACYT 60060--J. AHA is also grateful to SNI.

\end{document}